\documentclass[11pt,draftcls,onecolumn]{IEEEtran}
\usepackage{etex}
\usepackage{hyperref}

\usepackage{cite}

\usepackage{soul}

\usepackage[pdftex]{graphicx}
\usepackage{epstopdf}
 \usepackage{tikz}

\usepackage{amssymb}
\usepackage[cmex10]{amsmath}
\usepackage{cases}

\usepackage{array}
\usepackage{multirow}
\usepackage{bigdelim}
\usepackage{mdwtab}
\usepackage{ctable}
\DeclareMathOperator*{\esssup}{ess\,sup}

\DeclareMathOperator*{\argmin}{arg\,min}

\newtheorem{theorem}{Theorem}

\newtheorem{proposition}{Proposition}
\newtheorem{corollary}{Corollary}

\newtheorem{definition}{Definition}[section]

\begin{document}


\title{Misspecified and Asymptotically Minimax Robust Quickest Change Diagnosis}

%
\author{Timothy~L.~Molloy
%
\thanks{The author was with QUT, Brisbane, QLD, 4000, Australia. He is now with the Department of Electrical and Electronic Engineering, University of Melbourne, Parkville, VIC, 3010, Australia (e-mail:tim.molloy@unimelb.edu.au)}
}

\newcommand\copyrighttext{%
  \footnotesize \textcopyright 2020 IEEE. Personal use of this material is permitted.
  Permission from IEEE must be obtained for all other uses, in any current or future
  media, including reprinting/republishing this material for advertising or promotional
  purposes, creating new collective works, for resale or redistribution to servers or
  lists, or reuse of any copyrighted component of this work in other works.
  DOI: \href{https://doi.org/10.1109/TAC.2020.2985975}{10.1109/TAC.2020.2985975}}
\newcommand\copyrightnotice{%
\begin{tikzpicture}[remember picture,overlay]
\node[anchor=south,yshift=5pt] at (current page.south) {\fbox{\parbox{\dimexpr\textwidth-\fboxsep-\fboxrule\relax}{\copyrighttext}}};
\end{tikzpicture}%
}


\maketitle

\thispagestyle{empty}
\copyrightnotice

\begin{abstract}
\boldmath
The problem of quickly diagnosing an unknown change in a stochastic process is studied.
We establish novel bounds on the performance of misspecified diagnosis algorithms designed for changes that differ from those of the process, and pose and solve a new robust quickest change diagnosis problem in the asymptotic regime of few false alarms and false isolations.
Simulations suggest that our asymptotically robust solution offers a computationally efficient alternative to generalised likelihood ratio algorithms.
\end{abstract}


%
\IEEEpeerreviewmaketitle

\section{Introduction}

Quickest change diagnosis is concerned with simultaneously detecting and isolating  (or classifying) changes as one of $J$ possible change-types with minimal delay whilst avoiding false alarms and incorrect isolation decisions. 
Despite many applications in automatic control \cite{Hwang2010,Willsky1976,Basseville1988,James2018,Malladi1999,Lai1999,Molloy2019} and signal processing \cite{Oskiper2002,Nikiforov1995,Lai2000} requiring the diagnosis of unknown changes (e.g., fault detection and isolation), quickest change diagnosis with uncertain change-types has received limited attention \cite{Lai2000,Brodsky2008a}.
In this paper, we pose and solve a robust quickest change diagnosis problem with uncertain change-types, and we study the performance of misspecified diagnosis procedures that are designed for change-types that differ from those of the observed process.

Although the problem of detecting and isolating changes in signals and systems has a long history in applications such as fault diagnosis (cf.~ \cite{Willsky1976,Hwang2010,Malladi1999,Basseville1988}), the statistical decision problem of quickest change diagnosis was only recently posed in \cite{Nikiforov1995}.
Nikiforov \cite{Nikiforov1995} posed the problem of quickest change diagnosis by generalising the popular Lorden formulation of quickest change detection to penalise both detection and isolation delays under constraints on the mean times to false alarm and false isolation.
Several alternative formulations of quickest change diagnosis have since been proposed to address unpenalised false isolations that may occur for some change-types under the formulation of \cite{Nikiforov1995} (cf. \cite{Nikiforov2000a,Lai2000}).
Efficient recursive algorithms for solving these quickest change diagnosis problems have been found under the restrictive assumption that the possible post-change distributions (i.e., change-types) are known 
\cite{Oskiper2002,Nikiforov2000a,Nikiforov2003,Tartakovsky2014}.
For example, the quickest change diagnosis formulation of \cite{Nikiforov1995} is solved (in the asymptotic regime of few false alarms or isolations) by the matrix cumulative sum (MCUSUM) algorithm of \cite{Oskiper2002} when the change-types are known.

Existing quickest change diagnosis results are of limited applicability when the pre- or post-change distributions are unknown (i.e., the change-types are uncertain).
The treatments of Lai \cite{Lai2000}, and Brodsky and Darkhovsky \cite{Brodsky2008a} are notable exceptions and consider change-types involving distributions with unknown parameters.
For quickest change diagnosis problems with uncertain change-types, Lai \cite{Lai2000} proposes a mixture likelihood ratio (MLR) algorithm and a generalised likelihood ratio (GLR) algorithm whilst Brodsky and Darkhovsky \cite{Brodsky2008a} propose a modified GLR algorithm.
Lai \cite{Lai2000} makes limited progress in establishing the optimaility properties of MLR and GLR algorithms.
In contrast, Brodsky and Darkhovsky \cite{Brodsky2008a} prove the asymptotic optimality of their modified GLR algorithm under a new minimax-type criterion.
Unfortunately, the MLR, GLR and modified GLR algorithms are all computationally expensive due to the MLR algorithm involving integration over the set of unknown parameters, and the GLR and modified GLR algorithms involving constrained maximum likelihood estimation.

In contrast to the limited treatments of quickest change diagnosis with uncertain change-types, there has been considerable work in quickest change detection (without isolation) for cases where the pre and post-change distributions are uncertain but belong to known uncertainty sets of distributions \cite{Crow1994,Unnikrishnan2011a,Molloy2017c,Molloy2017d,Molloy2019}.
The rules that solve these robust quickest change detection problems minimise the worst case (i.e.~maximum) detection delays over the uncertainty sets of possible distributions, and have been observed to outperform more computationally complex GLR rules \cite{Unnikrishnan2011a,Molloy2017c,Molloy2017d}.
Motivated by robust quickest change detection, in this paper we seek to pose and solve a robust quickest change diagnosis problem with uncertain change-types.

The key contribution of this paper is the proposal and asymptotic solution of a robust quickest change diagnosis problem, and, the development of novel bounds on the performance of misspecified MCUSUM algorithms (i.e., MCUSUM algorithms designed with distributions that may differ from those of the observed process).
Through simulations, we illustrate that asymptotically robust algorithms can offer similar performance to GLR-type algorithms with a significant reduction in computational effort.
In contrast to the work of \cite{Lai2000} and \cite{Brodsky2008a}, our robust problem handles uncertainty in both pre- and post-change distributions under a formulation consistent with that of \cite{Nikiforov1995}.
Our work generalises and extends the robust quickest change detection (without isolation) treatments of \cite{Unnikrishnan2011a} and \cite{Molloy2017c} to settings involving the simultaneous detection and isolation of an unknown change from one of multiple possible post-change uncertainty sets.

The rest of this paper is structured as follows.
In Section \ref{sec:problem}, we pose our robust and misspecified problems.
In Section \ref{sec:misspecified}, we establish bounds on the performance of misspecified MCUSUM algorithms.
In Section \ref{sec:robust}, we present our main robustness results.
Finally, in Section \ref{sec:results} we illustrate our results in simulation and present conclusions in Section \ref{sec:conclusion}.

\section{Problem Statement}
\label{sec:problem}
Consider a sequence of random variables $Y_k$ for $k \geq 1$ each taking values in the set $\mathcal{Y} \subset \mathbb{R}^N$, and let $\mathcal{P}$ denote the set of all probability distributions on $\mathcal{Y}$.
Let $\lambda \geq 1$ be an unknown deterministic change-time such that $Y_k$ for $1 \leq k < \lambda$ are independent and identically distributed (i.i.d.) with marginal probability distribution $\nu_0 \in \mathcal{P}$, and $Y_k$ for $k \geq \lambda$  are i.i.d. with one of $J \geq 1$ possible marginal probability distributions $\nu_j \in \mathcal{P}$ where $1 \leq j \leq J$.
Let $P_\lambda^{{\nu_0},\nu_j}$ denote the probability law describing a $j$-type change (i.e., a change in distribution from $\nu_0$ to $\nu_j$) at time $\lambda \geq 1$ for $1 \leq j \leq J$, and let $E_\lambda^{{\nu_0},\nu_j} \left[\cdot\right]$ denote the expectation under $P_\lambda^{{\nu_0},\nu_j}$.
Similarly, let $P^{\nu_j}$ and $E^{\nu_j} [ \cdot ]$ denote the probability law and expectation, respectively, when $Y_k$ is i.i.d. with distribution $\nu_j$ for all for all $k \geq 1$.
Finally, let $\mathcal{F}_k \triangleq \sigma \left( Y_1, Y_2, \ldots, Y_k \right)$ denote the filtration generated by $Y_1, Y_2, \ldots, Y_k$.

In the problem of quickest change diagnosis, we observe $Y_k$ sequentially without knowledge of the change-time $\lambda$ or which of the $J$ post-change distributions $\nu_j$ the process $Y_k$ follows for $k \geq \lambda$.
Our aim is to simultaneously detect and isolate the change (i.e., diagnose the change-type $j$) as soon as possible after the change occurs whilst avoiding false alarms and incorrect isolation decisions.
The design of quickest diagnosis procedures thus involves formulating a suitable trade-off between detection and isolation delays, and the occurrence of false alarms and incorrect isolation decisions.

We shall follow the original formulation of \cite{Nikiforov1995} and characterise a quickest change diagnosis procedure $\eta \triangleq \left( T, d \right)$ by its stopping time $T$ and its isolation decision $d \in \mathcal{D}$ where $\mathcal{D} \triangleq \left\{ 1, 2, \ldots, J \right\}$ denotes the set of possible change-type decisions.
Both the stopping time $T$ and isolation decision $d$ are random variables (measurable) with respect to the filtration $\mathcal{F}_k$ for $k \geq 1$.
The maximum worst-average detection and isolation delay of a $j$-type change with $\eta$ is
\begin{align*}
 W \left( \eta, \nu \right)
 \triangleq \max_{1 \leq j \leq J} \sup_{\lambda \geq 1} \esssup E_\lambda^{{\nu_0},\nu_j} \left[ \left. \left( T - \lambda + 1 \right)^+ \right| \mathcal{F}_{\lambda-1} \right]
\end{align*}
where $x^+ \triangleq \max \left\{ x,0\right\}$ and $\nu \triangleq \{ \nu_i \in \mathcal{P} : 0 \leq i \leq J\}$.
Let us also define $\eta_t \triangleq \left( T_t, d_t \right)$ for $t \geq 1$ as an i.i.d. sequence of copies of $\eta$ where $\eta_t$ is the version of $\eta$ applied to $Y_{T_{t-1} + 1}, Y_{T_{t-1} + 2}, \ldots, Y_{T_{t}}$ with $T_0 = 0$.
The mean time to false alarm or false isolation of a $j$-type change with $\eta$ is then defined as the expectation
$
 E^{\nu_i} \left[ T_{d = j} \right]
$
where
\begin{align}
 \label{eq:isolationDelayDefinition}
 T_{d = j}
 &\triangleq \inf_{t \geq 1} \left\{ T_t : d_t = j\right\}
\end{align}
is the time that the first $j$-type change is detected by a copy of $\eta$ for $0 \leq i \leq J$ and $1 \leq j \leq J$ with $i \neq j$.
The quickest change diagnosis problem is then \cite{Nikiforov1995}
\begin{align}
 \label{eq:standard}
 \inf_{\eta \in C_\gamma \left( \nu \right)} W \left( \eta, \nu \right)
\end{align}
where $C_\gamma \left( \nu \right)$ is the set of diagnosis procedures satisfying the mean time to false alarm or false isolation constraint
\begin{align*}
 F (\eta, \nu)
 &\triangleq \min_{0 \leq i \leq J} \min_{\substack{1 \leq j \leq J\\ j \neq i}} E^{\nu_i} \left[ T_{d = j} \right] \geq \gamma
\end{align*}
for a given constant $1 < \gamma < \infty$.

Posing and solving \eqref{eq:standard} is frequently complicated in practice by unknown or uncertain pre- and post-change distributions. 
Similarly, in practice multiple distributions may describe a common change-type in the sense that a $j$-type change may be defined as a change from any pre-change distribution $\nu_0$ in the set $\mathcal{P}_0 \subset \mathcal{P}$ to any distribution $\nu_j$ belonging to one of $J \geq 1$ possible sets of post-change distributions $\mathcal{P}_j \subset \mathcal{P}$ for $1 \leq j \leq J$.
In this paper, we shall suppose that the distributions $\nu_j$ are unknown, but known to belong to the (mutually disjoint) uncertainty sets $\mathcal{P}_j \subset \mathcal{P}$ for $0 \leq j \leq J$.
We define the (known) uncertainty set of possible distribution collections as $\mathcal{Q} \triangleq \mathcal{P}_0 \times \mathcal{P}_1 \times \cdots \times \mathcal{P}_J$ such that $\nu = \{\nu_i \in \mathcal{P}_i : 0 \leq i \leq J\} \in \mathcal{Q}$.
We then pose robust quickest change diagnosis as the optimisation problem
\begin{equation}
 \label{eq:robust}
 \inf_{\eta \in C_\gamma}  \sup_{\nu \in \mathcal{Q}} W \left( \eta, \nu \right)
\end{equation}
where $C_\gamma$ is the set of procedures with respect to $\mathcal{F}_k$ satisfying the mean time to false alarm or false isolation constraint
\begin{equation*}
  \inf_{\nu \in \mathcal{Q}} F (\eta, \nu) \geq \gamma
\end{equation*}
for a given constant $1 < \gamma < \infty$.
Solutions to our robust quickest change diagnosis problem \eqref{eq:robust} have the minimax property of minimising the maximum (i.e., worst) detection and isolation delay over all possible pre-change and post-change distributions from the uncertainty sets $\{ \mathcal{P}_j \subset \mathcal{P} : 0 \leq j \leq J \}$.
In this paper, we shall specifically focus on identifying rules that solve our robust problem \eqref{eq:robust} in the asymptotic regime of few false alarms and few false isolations (i.e., as $\gamma \to \infty$).

In practice, we may also be unable to specify the uncertainty sets $\mathcal{P}_j$ to pose and solve \eqref{eq:robust}.
In these situations, a potentially naive approach is to apply a misspecified procedure that has optimality properties under \eqref{eq:standard} specified with nominal distributions that may differ from the true (unknown) distributions $\nu = \{ \nu_j : 0 \leq j \leq J \}$ of $Y_k$.
In this paper, we also seek to characterise the performance of misspecified procedures.

\section{Misspecified Quickest Change Diagnosis}
\label{sec:misspecified}
In this section, we revisit the Matrix Cumulative Sum (MCUSUM) algorithm of \cite{Oskiper2002} and examine its misspecified performance (i.e., its performance when it is designed with distributions that differ from those of the sequence $Y_k$).
We shall later exploit these misspecified results to identify asymptotic solutions to our robust problem \eqref{eq:robust}.

\subsection{Matrix Cumulative Sum (MCUSUM) Algorithm}

To revisit the MCUSUM algorithm and examine its misspecified performance, let us introduce the ordered pairs of distributions
\begin{align*}
 v_{ij} &\triangleq (v_{ij}^0, v_{ij}^1) \in \mathcal{P} \times \mathcal{P}
\end{align*}
for $0 \leq i \leq J$ and $1 \leq j \leq J$ with $i \neq j$, and let us collect these pairs in the set
\begin{align*}
\Upsilon
&\triangleq \left\{ v_{ij} : 0 \leq i,j \leq J, \; i \neq j, \; j \neq 0 \right\}.
\end{align*}
The MCUSUM algorithm $\hat{\eta} \left( \Upsilon \right) \triangleq (\hat{\tau}\left( \Upsilon \right), \hat{d} \left( \Upsilon \right))$ designed with pairs from $\Upsilon$ involves $J$ stopping rules of the form
\begin{align*}
 \hat{\tau}^j \left( \Upsilon \right)
 \triangleq \inf \left\{ k \geq 1 : \min_{\substack{0 \leq i \leq J \\ i \neq j}} S_k \left( v_{ij} \right) \geq h \right\},
\end{align*}
for detecting and isolating each of the possible $j$-type changes ($1 \leq j \leq J$).
Here, $h > 0$ is a threshold selected to control the time to false alarm or false isolation, and $S_k (v_{ij})$ for $0 \leq i \leq J$ and $1 \leq j \leq J$ with $i \neq j$ are the cumulative sum (CUSUM) test statistics
\begin{align*}
 S_k (v_{ij})
 &\triangleq
    \max_{1 \leq n \leq k} Z_n^k (v_{ij})
\end{align*}
for $k \geq 1$ where
\begin{align}
  \label{eq:llrs}
  Z_n^k (v_{ij})
  &\triangleq \sum_{\ell = n}^k \log \dfrac{\mathrm{d} v_{ij}^1}{\mathrm{d } v_{ij}^0} \left( Y_\ell \right)
 \end{align}
is the log-likelihood ratio between the distributions $v_{ij}^0$ and $v_{ij}^1$, and $\frac{\mathrm{d} v_{ij}^1}{\mathrm{d } v_{ij}^0} (\cdot)$ denotes the Radon-Nikodym derivative of $v_{ij}^1$ with respect to $v_{ij}^0$.
The stopping rule $\hat{\tau} \left( \Upsilon \right)$ of the MCUSUM algorithm $\hat{\eta} \left( \Upsilon \right)$ is the first of the $J$ stopping rules, namely,
\begin{align}
 \label{eq:mcusumStoppingRule}
 \hat{\tau}\left( \Upsilon \right)
 &\triangleq \min_{1 \leq j \leq J} \hat{\tau}^j\left( \Upsilon \right),
\end{align}
and the MCUSUM isolation decision $\hat{d}(\Upsilon)$ is
\begin{align}
 \label{eq:mcusumDecision}
 \hat{d} \left( \Upsilon \right)
 &\triangleq \argmin_{1 \leq j \leq J} \hat{\tau}^j\left( \Upsilon \right).
\end{align}

The MCUSUM algorithm $\hat{\eta} \left( \Upsilon \right)$ has an efficient recursive implementation since its test statistics $S_k(v_{ij})$ are calculable via the recursions
$S_k (v_{ij}) = S_{k-1}^+ (v_{ij}) + Z_k^k (v_{ij})$ for $k \geq 1$ with $S_0 (v_{ij}) \triangleq 0$.
Loosely speaking, the statistics $S_k(v_{0j})$ are used in the rules $\hat{\tau}^j \left( \Upsilon \right)$ to test a null hypothesis that no-change has occurred against an alternative hypothesis that a $j$-type change has occurred.
Similarly, the statistics $S_k(v_{ij})$ for $1 \leq i,j \leq J$ with $i \neq j$, are used to test a null hypothesis that an $i$-type change has occurred against an alternative hypothesis that a $j$-type change has occurred.

In \cite{Oskiper2002}, the asymptotic optimality of the MCUSUM algorithm $\hat{\eta} \left( \Upsilon^* \right)$ under \eqref{eq:standard} is established when the set of pairs $\Upsilon^*$ matches the distributions $\nu$ of $Y_k$ in the sense that
\begin{equation*}
\Upsilon^*
\triangleq \left\{ v_{ij} = v_{ij}^* \triangleq (\nu_i,\nu_j) : 0 \leq i,j \leq J, \; i \neq j, \; j \neq 0 \right\}.
\end{equation*}
Here, we shall examine the performance of the MCUSUM algorithm $\hat{\eta} \left( \bar{\Upsilon} \right)$ when it is designed with a set of pairs 
\begin{equation*}
\bar{\Upsilon}
\triangleq \left\{ v_{ij} = \bar{v}_{ij} \triangleq (\bar{v}_{ij}^0,\bar{v}_{ij}^1) : 0 \leq i,j \leq J, \; i \neq j, \; j \neq 0 \right\}
\end{equation*}
that may differ from the set $\Upsilon^*$ associated with the true distributions $\nu$ of the sequence $Y_k$.
Under our construction of the potentially misspecified MCUSUM algorithm $\hat{\eta} \left( \bar{\Upsilon} \right)$, there is no requirement for any of the pairs $\bar{v}_{ij} = (\bar{v}_{ij}^0,\bar{v}_{ij}^1) \in \bar{\Upsilon}$ to share the same distribution for testing a specific change-type.
For example, the no-change distribution $\bar{v}_{0j}^0$ used in $S_k(\bar{v}_{0j})$ to test for the occurrence of a $j$-type change may be different to the no-change distribution $\bar{v}_{0i}^0$ used in $S_k(\bar{v}_{0i})$ to test for the occurrence of an $i$-type change, $i \neq j$.
To analyse the performance of the potentially misspecified MCUSUM algorithm $\hat{\eta} \left( \bar{\Upsilon} \right)$, we require the concept of relative entropy.

\subsection{Relative Entropy and Likelihood Ratio Convergence}
Consider any two distributions $\mu,\bar{\mu} \in \mathcal{P}$.
The relative entropy of $\bar{\mu}$ from $\mu$ is defined as \cite[p. 26]{Dupuis1997}
\begin{align*}
 D(\mu \| \bar{\mu})
 &\triangleq \begin{cases}
              {\displaystyle \int_{\mathcal{Y}}} \log \dfrac{\mathrm{d} \mu}{\mathrm{d } \bar{\mu}} (Y) \; \mathrm{d } \mu &\text{if } \mu \ll \bar{\mu}, \\
              + \infty &\text{otherwise}
             \end{cases}
\end{align*}
where $\mu \ll \bar{\mu}$ denotes that $\mu$ is absolutely continuous with $\bar{\mu}$.
The relative entropy 
has an (informal) interpretation as a pseudo-distance between the distributions $\mu$ and $\bar{\mu}$ because it is non-negative with $D ( \mu \| \bar{\mu} ) = 0$ if and only if $\mu = \bar{\mu}$ \cite[Lemma 1.4.1]{Dupuis1997}.
We shall use the relative entropy to describe the asymptotic behaviour of the log-likelihood ratios \eqref{eq:llrs} as $k \to \infty$.
Specifically, consider the distributions $\nu_0, \nu_j \in \mathcal{P}$ and the pair $\bar{v}_{ij} \in \bar{\Upsilon}$ and suppose that $D \left( \nu_j \left\| \bar{v}_{ij}^1 \right. \right) < \infty$ and $D \left( \nu_j \left\| \bar{v}_{ij}^0 \right. \right) < \infty$.
The weak law of large numbers gives that
\begin{align}
 \label{eq:weakLaw}
 \lim_{k \to \infty} P_\lambda^{\nu_0,\nu_j} \left( \left| k^{-1} Z_n^{n + k - 1} (\bar{v}_{ij}) - \Delta_{ij}^{\nu,\bar{v}} \right| \geq \epsilon \right) = 0
 \end{align}
for all $\epsilon > 0$, all $\lambda \geq 1$ and all $n \geq \lambda$ where we define
\begin{align*}
 \Delta_{ij}^{\nu,\bar{v}}
 \triangleq D \left( \nu_j \left\| \bar{v}_{ij}^0 \right. \right) - D \left( \nu_j \left\| \bar{v}_{ij}^1 \right. \right)
\end{align*}
and note that $ E_\lambda^{\nu_0,\nu_j} \left[ Z_n^n (\bar{v}_{ij}) \right]
= \Delta_{ij}^{\nu,\bar{v}}
$
for all $\lambda \geq 1$ and all $n \geq \lambda$ due to properties of the logarithm, and the definitions of the relative entropy and the log-likelihood ratio \eqref{eq:llrs}.

\subsection{MCUSUM Performance Bounds}

We first provide an upper bound on the delays of the potentially misspecified MCUSUM algorithm $\hat{\eta}(\bar{\Upsilon})$.

\begin{theorem}
 \label{theorem:delayUpperBound}
 Consider the set of pairs $\bar{\Upsilon}$ and the distributions $\nu =\{ \nu_j \in \mathcal{P} : 0 \leq j \leq J\}$.
 If $0 < \Delta_{ij}^{\nu,\bar{v}} < \infty$ for all $0 \leq i \leq J$ and all $1 \leq j \leq J$ with $i \neq j$, then the MCUSUM algorithm $\hat{\eta} \left( \bar{\Upsilon} \right)$ satisfies
 \begin{align}
  \label{eq:delayUpperBound}
  W \left( \hat{\eta} \left( \bar{\Upsilon} \right), \nu \right)
  &\leq (1 + o(1)) \left( \dfrac{h}{\Delta_*^{\nu,\bar{v}}} \right)
 \end{align}
 as $h \to \infty$ where
 \begin{align*}
 \Delta_*^{\nu,\bar{v}}
 \triangleq \min_{0 \leq i \leq J} \min_{\substack{1 \leq j \leq J\\j \neq i}} \Delta_{ij}^{\nu,\bar{v}}.
\end{align*}
\end{theorem}
\begin{IEEEproof}
Consider any $1 \leq j \leq J$ and any arbitrary $\epsilon \in \left( 0, \min \{ 1, \Delta_*^{\nu,\bar{v}} \} \right)$.
The theorem condition that $0 < \Delta_{ij}^{\nu,\bar{v}} < \infty$ for $0 \leq i \leq J$ with $i \neq j$ implies that \eqref{eq:weakLaw} holds, and so
\begin{align*}
\lim_{k \to \infty} P_\lambda^{\nu_0,\nu_j} \left( k^{-1} Z_n^{n + k - 1} (\bar{v}_{ij}) \leq (1 - \epsilon)\Delta_{ij}^{\nu,\bar{v}} \right) = 0
\end{align*}
for all $\lambda \geq 1$, all $n \geq \lambda$, and all $0 \leq i \leq J$ with $i \neq j$.
Since $\Delta_{ij}^{\nu,\bar{v}} \geq \Delta_*^{\nu,\bar{v}}$, we have that
\begin{align*}
\lim_{k \to \infty} P_\lambda^{\nu_0,\nu_j} \left( \min_{\substack{0 \leq i \leq J\\i \neq j}} Z_n^{n + k - 1} (\bar{v}_{ij}) \leq k (1 - \epsilon)\Delta_*^{\nu,\bar{v}} \right) = 0
\end{align*}
for all $\lambda \geq 1$ and all $n \geq \lambda$.
By defining $k_\epsilon$ as the largest integer less than $h\left((1 - \epsilon)\Delta_*^{\nu,\bar{v}}\right)^{-1}$ we then have that 
\begin{align*}
 \lim_{h \to \infty} P_{\lambda}^{\nu_0,\nu_j} \left( \min_{\substack{0 \leq i \leq J\\i \neq j}} Z_{n}^{n + k_\epsilon - 1} (\bar{v}_{ij})
 < h \right) 
 = 0
\end{align*}
for all $\lambda \geq 1$ and $n \geq \lambda$.
Thus, for any arbitrary $\epsilon \in \left( 0, \min \{ 1, \Delta_*^{\nu,\bar{v}} \}\right)$ and sufficiently large $h$, we have that
\begin{align}
 \label{eq:secondProbBound}
 P_{\lambda}^{\nu_0,\nu_j} \left( \min_{\substack{0 \leq i \leq J\\i \neq j}} Z_{n}^{n + k_\epsilon - 1} (\bar{v}_{ij})
 < h \right) 
 < \epsilon
\end{align}
for all $\lambda \geq 1$ and $n \geq \lambda$.

Now, by recalling the definition of the $j$-type stopping rules $\hat{\tau}^j \left( \bar{\Upsilon} \right)$ we have that
 \begin{align*}
   &\esssup P_{\lambda}^{\nu_0,\nu_j} \left( \left. \hat{\tau}^j \left( \bar{\Upsilon} \right) - \lambda + 1 > t k_\epsilon \right| \mathcal{F}_{\lambda - 1} \right)\\
   &\leq P_{\lambda}^{\nu_0,\nu_j} \left( \min_{\substack{0 \leq i \leq J\\i \neq j}}  Z_{\lambda + (\ell - 1)k_\epsilon}^{\lambda + \ell k_\epsilon - 1} (\bar{v}_{ij}) < h \text{ for } 1 \leq \ell \leq t \right)\\
   &= \prod_{\ell = 1}^t P_{\lambda}^{\nu_0,\nu_j} \left( \min_{\substack{0 \leq i \leq J\\i \neq j}}  Z_{\lambda + (\ell - 1)k_\epsilon}^{\lambda + \ell k_\epsilon - 1} (\bar{v}_{ij}) < h \right)
 \end{align*}
 for all $\lambda \geq 1$, and all $t \geq 1$.
 Applying \eqref{eq:secondProbBound} then gives that 
\begin{align}
 \label{eq:powerProbBound}
 \esssup P_{\lambda}^{\nu_0,\nu_j} \left( \left. \hat{\tau}^j \left( \bar{\Upsilon} \right) - \lambda + 1 > t k_\epsilon \right| \mathcal{F}_{\lambda - 1} \right)
 &\leq \epsilon^t
\end{align}
for all $\lambda \geq 1$, all $t \geq 1$, and sufficiently large $h$.
The definition of the MCUSUM rule \eqref{eq:mcusumStoppingRule} then implies that
\begin{align*}
 &\esssup E_{\lambda}^{\nu_0,\nu_j} \left[ \left. \left( \hat{\tau} \left( \bar{\Upsilon} \right) - \lambda + 1\right)^+ \right| \mathcal{F}_{\lambda-1} \right]\\
 &\leq \esssup E_{\lambda}^{\nu_0,\nu_j} \left[ \left. \left( \hat{\tau}^j \left( \bar{\Upsilon} \right) - \lambda + 1\right)^+ \right| \mathcal{F}_{\lambda-1} \right]\\
 &= \esssup \int_0^\infty P_\lambda^{\nu_0,\nu_j} \left( \left. \left( \hat{\tau}^j \left( \bar{\Upsilon} \right) - \lambda + 1 \right)^+ > y \right| \mathcal{F}_{\lambda-1} \right) \, \text{d}y\\
 &\leq k_\epsilon \sum_{t = 0}^{\infty} \epsilon^t 
 = k_\epsilon \dfrac{1}{1 - \epsilon}
\end{align*}
for all $1 \leq j \leq J$, all $\lambda \geq 1$, and sufficiently large $h$ where the last inequality follows by bounding the integral by the sum of rectangles and \eqref{eq:powerProbBound}, and the last equality follows from the sum of a geometric series since $\epsilon \in \left( 0, \min \{ 1, \Delta_*^{\nu,\bar{v}} \} \right) < 1$.
The theorem assertion follows by recalling the definition of $k_\epsilon$ and since $\epsilon$ can be arbitrarily close to $0$.
\end{IEEEproof}

We next establish a bound on the mean time to false alarm or false isolation of $\hat{\eta} \left( \bar{\Upsilon} \right)$.

\begin{theorem}
 \label{theorem:falseBound}
  Consider the set of pairs $\bar{\Upsilon}$ and the distributions $\nu =\{ \nu_i \in \mathcal{P} : 0 \leq i \leq J\}$, and suppose that
 \begin{align}
  \label{eq:likelihoodRatioBound}
    E^{\nu_i} \left[ \dfrac{\mathrm{d} \bar{v}_{ij}^1}{\mathrm{d } \bar{v}_{ij}^0} (Y) \right] \leq 1
 \end{align}
 for all $0 \leq i \leq J$ and all $1 \leq j \leq J$ with $i \neq j$.
 Then the mean time to false alarm or false isolation of $\hat{\eta} \left( \bar{\Upsilon} \right)$ satisfies
 \begin{align*}
  F \left( \hat{\eta} \left( \bar{\Upsilon} \right), \nu \right)
  = \min_{0 \leq i \leq J} \min_{\substack{1 \leq j \leq J\\ j \neq i}} E^{\nu_i} \left[  \hat{\tau}_{\hat{d} = j} \left( \bar{\Upsilon} \right) \right]
  &\geq e^{h}
 \end{align*}
 for any $h > 0$ where
 \begin{align*}
 \hat{\tau}_{\hat{d} = j} \left( \bar{\Upsilon} \right)
 &\triangleq \inf_{t \geq 1} \left\{ \hat{\tau}_t : \hat{d}_t = j\right\}
\end{align*}
is the first time that a copy $(\hat{\tau}_t, \hat{d}_t)$ of $\hat{\eta} \left( \bar{\Upsilon} \right)$ applied to $Y_{\hat{\tau}_{t-1} + 1}, Y_{\hat{\tau}_{t-1} + 2}, \ldots, Y_{\hat{\tau}_{t}}$ for $t \geq 1$ with $\hat{\tau}_0 = 0$ detects a $j$-type change.
\end{theorem}
\begin{IEEEproof}
The definition of $\hat{\eta} \left( \bar{\Upsilon} \right)$ and the construction of the first time to a $j$-type detection in \eqref{eq:isolationDelayDefinition} implies that the stopping time $\hat{\tau}_t$ applied to $Y_{\hat{\tau}_{t-1} + 1}, Y_{\hat{\tau}_{t-1} + 2}, \ldots, Y_{\hat{\tau}_{t}}$ for $t \geq 1$ with $\hat{\tau}_0 = 0$ is
 $
  \hat{\tau}_t
  = \min \left\{ \hat{\tau}_t^1, \hat{\tau}_t^2, \ldots, \hat{\tau}_t^J \right\}
 $
 where
 \begin{align*}
  \hat{\tau}_t^j
  &\triangleq \inf\left\{ k \geq \hat{\tau}_{t-1} + 1 : \min_{\substack{0 \leq i \leq J \\ i \neq j}} S_{t,k} (\bar{v}_{ij}) \geq h \right\}
 \end{align*}
 with $\hat{\tau}_0^j = 0$ for $1 \leq j \leq J$ and 
 \begin{align*}
  S_{t,k} (\bar{v}_{ij})
  &\triangleq \max_{\hat{\tau}_{t-1} < n \leq k} Z_n^k (\bar{v}_{ij})
\end{align*}
for all $0 \leq i \leq J$ and all $1 \leq j \leq J$, $i \neq j$.
Here, $S_{t,k} (\bar{v}_{ij}) \triangleq - \infty$ for $k < \hat{\tau}_{t-1} + 1$.
By noting the different maximisation ranges in the test statistics $S_{t,k} (\bar{v}_{ij})$ and $S_{1,k} (\bar{v}_{ij})$, we have
\begin{align*}
\min_{\substack{0 \leq i \leq J \\ i \neq j}} S_{t,k} (\bar{v}_{ij})
&\leq  \min_{\substack{0 \leq i \leq J \\ i \neq j}} S_{1,k} (\bar{v}_{ij})
\leq S_k \left(  \bar{v}_{ij} \right)
\end{align*}
for all $t \geq 1$, all $k \geq 1$, all $0 \leq i \leq J$ and all $1 \leq j \leq J$ with $i \neq j$.
Thus,
$
 \hat{\tau}_t^j
 \geq \hat{\tau}_1^j
 \geq \tau_C (\bar{v}_{ij})
$ for all $t \geq 1$, all $0 \leq i \leq J$ and all $1 \leq j \leq J$ with $i \neq j$ where we define $\tau_C (\bar{v}_{ij})$ as the CUSUM stopping rule 
\begin{equation*}
  \tau_C \left( \bar{v}_{ij} \right)
  \triangleq \inf \left\{ k \geq 1 : S_k \left(  \bar{v}_{ij} \right) \geq h \right\}.
\end{equation*}
Hence, if $\hat{\tau}_{\hat{d}  = j} \left( \bar{\Upsilon} \right)  = \hat{\tau}_t$ for any $t \geq 1$, then
 $ \hat{\tau}_{\hat{d}  = j} \left( \bar{\Upsilon} \right)  
 = \hat{\tau}_t^j \geq \tau_C (\bar{v}_{ij})$
 and so
 $
  E^{\nu_i} [ \hat{\tau}_{\hat{d}  = j} \left( \bar{\Upsilon} \right) ]
  \geq  E^{\nu_i} \left[ \tau_C (\bar{v}_{ij}) \right]
 $
for all $0 \leq i \leq J$ and all $1 \leq j \leq J$ with $i \neq j$.
The theorem assertion follows since the misspecified CUSUM false alarm result of \cite[Theorem 2]{Molloy2017c} holding under \eqref{eq:likelihoodRatioBound} gives that
$
E^{\nu_i} \left[ \tau_C (\bar{v}_{ij}) \right] \geq e^h
$ for all $h > 0$, $0 \leq i \leq J$, and $1 \leq j \leq J$ with $i \neq j$.
\end{IEEEproof}

The conditions of Theorems \ref{theorem:delayUpperBound} and \ref{theorem:falseBound} are sufficient for avoiding extreme instances of misspecification in the MCUSUM algorithm $\hat{\eta} \left( \bar{\Upsilon} \right)$.
Indeed, to correctly detect and isolate a $j$-type change, the MCUSUM algorithm $\hat{\eta} \left( \bar{\Upsilon} \right)$ relies on its $j$-type rule $\hat{\tau}^j \left( \bar{\Upsilon} \right)$ stopping first after a change under $P_\lambda^{\nu_0,\nu_j}$.
If $\Delta_{ij}^{\nu,\bar{v}} < 0$, then $\hat{\tau}^j \left( \bar{\Upsilon} \right)$ may be delayed after a $j$-type change since the statistics $S_k \left( \bar{v}_{0j} \right)$ will remain near zero after a change under $P_\lambda^{\nu_0,\nu_j}$ and will be very slow in exceeding the threshold $h$.
If $E^{\nu_i} \left[ \frac{\mathrm{d} \bar{v}_{ij}^1}{\mathrm{d } \bar{v}_{ij}^0} (Y) \right] > 1$, then $\hat{\tau}^j \left( \bar{\Upsilon} \right)$ may stop quickly without a $j$-type change since the statistics $S_k (\bar{v}_{ij})$ may grow under $P^{\nu_i}$ causing false alarms when $i = 0$ or false isolations when $j \neq i \neq 0$.
In particular, the MCUSUM algorithm may have a detection delay longer than its mean time to false alarm or false isolation when $E^{\nu_i} \left[ \frac{\mathrm{d} \bar{v}_{ij}^1}{\mathrm{d } \bar{v}_{ij}^0} (Y) \right] > 1$ (see \cite[Lemma 4.3]{Molloy2015a} for details of this phenomena when $J = 1$).
The MCUSUM algorithm is thus unlikely to perform well when the conditions of Theorems 1 and 2 are violated.

By recalling the pseudo-distance interpretation of relative entropy, the condition that $\Delta_{ij}^{\nu,\bar{v}} > 0$ in Theorem \ref{theorem:delayUpperBound} is intuitively satisfied when the true post-change (or alternative) distribution $\nu_j$ is closer to the misspecified post-change (or alternative) distribution $\bar{v}_{ij}^1$ than it is to the misspecified pre-change (or null) distribution $\bar{v}_{ij}^0$ in the sense that $D \left( \nu_j \left\| \bar{v}_{ij}^0 \right. \right) > D \left( \nu_j \left\| \bar{v}_{ij}^1 \right. \right)$.
Similarly, by taking the logarithm of \eqref{eq:likelihoodRatioBound} and apply Jensen's inequality, we see that a necessary condition for \eqref{eq:likelihoodRatioBound} to hold in Theorem \ref{theorem:falseBound} is that
\begin{align*}
    0 
    &\geq E^{\nu_i} \left[ \log \dfrac{\mathrm{d} \bar{v}_{ij}^1}{\mathrm{d } \bar{v}_{ij}^0} (Y) \right] = D \left( \nu_i \left\| \bar{v}_{ij}^0 \right. \right) - D \left( \nu_i \left\| \bar{v}_{ij}^1 \right. \right)
\end{align*}
which holds when the true pre-change (or null) distribution $\nu_i$ is closer (in a relative-entropy sense) to the misspecified pre-change (or null) distribution $\bar{v}_{ij}^0$ than to the misspecified post-change (or alternative) distribution $\bar{v}_{ij}^1$.
The conditions of Theorems \ref{theorem:delayUpperBound} and \ref{theorem:falseBound} thus intuitively require the misspecified distributions used in the MCUSUM algorithm to test each change-type to be amongst the closet distributions (in a relative-entropy sense) from the pairs in $\bar{\Upsilon}$ to the corresponding true change-type distribution from $\nu$.
We next discuss conditions on the sets $\{ \mathcal{P}_i : 0 \leq i \leq J \}$ such that all distributions in them satisfy the conditions of Theorems \ref{theorem:delayUpperBound} and \ref{theorem:falseBound}, and so that we may identify asymptotic solutions to our robust problem \eqref{eq:robust}. 

\section{Minimax Robust Quickest Change Diagnosis}
\label{sec:robust}
The solution of our robust problem \eqref{eq:robust} is simplified when there exist distributions from the uncertainty sets $\{ \mathcal{P}_i : 0 \leq i \leq J \}$ such that the algorithm that solves our robust problem \eqref{eq:robust} is the algorithm that also solves the non-robust problem \eqref{eq:standard} specified by these least favourable distributions (LFDs).
In this section, we will identify LFDs and an asymptotic solution to our robust problem under a novel dual stochastic boundedness condition on the uncertainty sets.

\subsection{Stochastically Bounded Uncertainty Sets}
Dual stochastic boundedness is defined as follows.

\begin{definition}[Dual Stochastic Boundedness]
\label{def:DSB}
We say that the sets $\{ \mathcal{P}_i \subset \mathcal{P} : 0 \leq i \leq J\}$ are dually stochastically bounded by the distributions $\tilde{\nu} \triangleq \{ \tilde{\nu}_i \in \mathcal{P}_i : 0 \leq i \leq J\} \in \mathcal{Q}$ and the pairs $\hat{v}_{ij} \triangleq (\hat{v}_{ij}^0,\hat{v}_{ij}^1) \in \mathcal{P}_i \times \mathcal{P}_j$ from the set 
\begin{equation*}
\hat{\Upsilon}
\triangleq \left\{ v_{ij} = \hat{v}_{ij} : 0 \leq i,j \leq J, \; i \neq j, \; j \neq 0 \right\}
\end{equation*}
 when
\begin{align}
 \label{eq:DSB1}
 \begin{split}
 &\min_{0 \leq i \leq J} \min_{\substack{1 \leq j \leq J\\j \neq i}} \left[ D \left( \tilde{\nu}_j \left\| \hat{v}_{ij}^0 \right. \right) - D \left( \tilde{\nu}_j \left\| \hat{v}_{ij}^1 \right. \right) \right]\\
 &\quad\leq\min_{0 \leq i \leq J} \min_{\substack{1 \leq j \leq J\\j \neq i}} \inf_{\nu_j \in \mathcal{P}_j} \left[ D \left( \nu_j \left\| \hat{v}_{ij}^0 \right. \right) - D \left( \nu_j \left\| \hat{v}_{ij}^1 \right. \right) \right]
 \end{split}
\end{align}
holds together with
\begin{align}
 \label{eq:DSB2}
 \begin{split}
 &\min_{0 \leq i \leq J} \min_{\substack{1 \leq j \leq J\\j \neq i}} D \left( \tilde{\nu}_j \left\| \tilde{\nu}_i \right. \right)\\
 &\quad \leq \min_{0 \leq i \leq J} \min_{\substack{1 \leq j \leq J\\j \neq i}} \left[ D \left( \tilde{\nu}_j \left\| \hat{v}_{ij}^0 \right. \right) - D \left( \tilde{\nu}_j \left\| \hat{v}_{ij}^1 \right. \right) \right],
 \end{split}
\end{align}
and
\begin{align}
 \label{eq:DSB3}
 \max_{0 \leq i \leq J} \max_{\substack{1 \leq j \leq J\\ j \neq i}} \sup_{\nu_i \in \mathcal{P}_i} E^{\nu_i} \left[ \dfrac{\mathrm{d} \hat{v}_{ij}^1}{\mathrm{d } \hat{v}_{ij}^0} (Y) \right] \leq 1.
\end{align}
\end{definition}

If the sets $\{ \mathcal{P}_i : 0 \leq i \leq J \}$ are dually stochastically bounded by $\tilde{\nu}$ and pairs from $\hat{\Upsilon}$, then the conditions of Theorems \ref{theorem:delayUpperBound} and \ref{theorem:falseBound} will hold for all $\nu \in \mathcal{Q}$ with $\bar{\Upsilon} = \hat{\Upsilon}$.
Dual stochastic boundedness also generalises and extends the following concept of weak stochastic boundedness previously used to identify LFDs for robust quickest change detection (without isolation) in \cite{Molloy2017c}.

\begin{definition}[Weak Stochastic Boundedness \cite{Molloy2017c}]
\label{definition:WSB}
The pair of uncertainty sets $\left( \mathcal{P}_i, \mathcal{P}_j \right)$ is said to be weakly stochastically bounded by the pair of distributions $\hat{v}_{ij} = (\hat{v}_{ij}^0,\hat{v}_{ij}^1) \in \mathcal{P}_i \times \mathcal{P}_j$ when
\begin{align}
 \label{eq:WSB1}
 D \left( \nu_j \left\| \hat{v}_{ij}^0 \right. \right) - D \left( \nu_j \left\| \hat{v}_{ij}^1 \right. \right)
 &\geq D \left( \hat{v}_{ij}^1 \left\| \hat{v}_{ij}^0 \right. \right)
\end{align}
for all $\nu_j \in \mathcal{P}_j$, and
\begin{align}
 \label{eq:WSB2}
 E^{\nu_i} \left[ \dfrac{\mathrm{d} \hat{v}_{ij}^1}{\mathrm{d } \hat{v}_{ij}^0} (Y) \right]
 \leq 1
\end{align}
for all $\nu_i \in \mathcal{P}_i$.
\end{definition}

Dual stochastic boundedness (Definition \ref{def:DSB}) is equivalent to weak stochastic boundedness (Definition \ref{definition:WSB}) when $J = 1$ and the distributions $\tilde{\nu}$ are taken as the pair $\hat{v}_{01} = (\tilde{\nu}_0,\tilde{\nu}_1)$.
In the following proposition, we show that weak stochastic boundedness of the pairs of uncertainty sets $\left( \mathcal{P}_i, \mathcal{P}_j \right)$ from $\{ \mathcal{P}_j \subset \mathcal{P} : 0 \leq j \leq J\}$ can also be used to establish dual stochastic boundedness when $J > 1$.

\begin{proposition}
 \label{proposition:weakImpliesDual}
 Consider the sets $\{ \mathcal{P}_j \subset \mathcal{P} : 0 \leq j \leq J\}$ and the distributions $\tilde{\nu} = \{ \tilde{\nu}_i \in \mathcal{P}_i : 0 \leq i \leq J \} \in \mathcal{Q}$.
 If the pairs of sets $\left( \mathcal{P}_i, \mathcal{P}_j \right)$ are each weakly stochastically bounded by the pairs $\hat{v}_{ij} \in \hat{\Upsilon}$ for all $0 \leq i \leq J$ and all $1 \leq j \leq J$ with $i \neq j$ in the sense of Definition \ref{definition:WSB}, and
 \begin{align}
 \label{eq:pDSB1}
 \begin{split}
 \min_{0 \leq i \leq J} \min_{\substack{1 \leq j \leq J\\j \neq i}} \inf_{\nu_j \in \mathcal{P}_j} \left[ D \left( \nu_j \left\| \hat{v}_{ij}^0 \right. \right) - D \left( \nu_j \left\| \hat{v}_{ij}^1 \right. \right) \right]
 &= \Delta_*^{\tilde{\nu},\hat{v}}
 \end{split}
\end{align}
holds together with
\begin{align}
    \label{eq:pDSB2}
    \min_{0 \leq i \leq J} \min_{\substack{1 \leq j \leq J\\j \neq i}} D \left( \tilde{\nu}_j \left\| \tilde{\nu}_i \right. \right)
    \leq
    \min_{0 \leq i \leq J} \min_{\substack{1 \leq j \leq J\\j \neq i}} D \left( \hat{v}_{ij}^1 \left\| \hat{v}_{ij}^0 \right. \right),
\end{align}
then the sets $\{ \mathcal{P}_j \subset \mathcal{P} : 0 \leq j \leq J\}$ are dually stochastically bounded by the distributions $\tilde{\nu}$ and the pairs from $\hat{\Upsilon}$.
 \end{proposition}
\begin{IEEEproof}
The weak stochastic boundedness of $\left( \mathcal{P}_i, \mathcal{P}_j \right)$ by $\hat{v}_{ij}$ for all $0 \leq i \leq J$ and all $1 \leq j \leq J$ with $i \neq j$ implies that \eqref{eq:WSB2} holds for all $0 \leq i \leq J$ and all $1 \leq j \leq J$ with $j \neq i$.
Hence, \eqref{eq:DSB3} holds.
Weak stochastic boundedness of $\left( \mathcal{P}_i, \mathcal{P}_j \right)$ by $\hat{v}_{ij}$ for all $0 \leq i \leq J$ and all $1 \leq j \leq J$ with $i \neq j$ also implies that \eqref{eq:WSB1}
holds for all $0 \leq i \leq J$ and all $1 \leq j \leq J$ with $j \neq i$.
Hence, 
\begin{align*}
 \min_{0 \leq i \leq J} \min_{\substack{1 \leq j \leq J\\j \neq i}} D \left( \hat{v}_{ij}^1 \left\| \hat{v}_{ij}^0 \right. \right)
 \leq \min_{0 \leq i \leq J} \min_{\substack{1 \leq j \leq J\\j \neq i}} \inf_{\nu_j \in \mathcal{P}_j} \Delta_{ij}^{\nu,\hat{v}}
 \leq \Delta_*^{\tilde{\nu},\hat{v}}
\end{align*}
where the last inequality follows from the definition of the infimum.
By applying \eqref{eq:pDSB2}, we have that 
\begin{align*}
    \min_{0 \leq i \leq J} \min_{\substack{1 \leq j \leq J\\j \neq i}} D \left( \tilde{\nu}_j \left\| \tilde{\nu}_i \right. \right)
    \leq \Delta_*^{\tilde{\nu},\hat{v}}
\end{align*}
and so \eqref{eq:DSB2} holds.
Finally, \eqref{eq:DSB1} can only hold with equality since the definition of the infimum implies that
\begin{align*}
    \min_{0 \leq i \leq J} \min_{\substack{1 \leq j \leq J\\j \neq i}} \inf_{\nu_j \in \mathcal{P}_j} \Delta_{ij}^{\nu,\hat{v}}
    \leq \Delta_*^{\tilde{\nu},\hat{v}}
\end{align*}
and so \eqref{eq:pDSB1} implies \eqref{eq:DSB1}.
The proof is complete.
\end{IEEEproof}

Proposition \ref{proposition:weakImpliesDual} implies that one procedure for showing that the uncertainty sets $\{ \mathcal{P}_j \subset \mathcal{P} : 0 \leq j \leq J\}$ are dually stochastically bounded is to first show that the pairs of sets $\left( \mathcal{P}_i, \mathcal{P}_j \right)$ are each weakly stochastically bounded by pairs of distributions $\hat{v}_{ij}$ from a set $\hat{\Upsilon}$.
Distributions $\tilde{\nu}$ such that \eqref{eq:pDSB1} and \eqref{eq:pDSB2} hold (if any exist) can then be found by directly solving the optimisations implied by \eqref{eq:pDSB1} and \eqref{eq:pDSB2}.
Indeed, candidate bounding distributions $\tilde{\nu}_j$ can be found by directly solving the optimisation in \eqref{eq:pDSB1}, whilst \eqref{eq:pDSB2} implies that the pair of distributions $\hat{v}_{ij}$ that minimise the relative entropy $D \left( \hat{v}_{ij}^1 \left\| \hat{v}_{ij}^0 \right. \right)$ are natural candidates for $(\tilde{\nu}_i,\tilde{\nu}_j)$.
If the distributions satisfying \eqref{eq:pDSB1} and \eqref{eq:pDSB2} do not contradict, then Proposition \ref{proposition:weakImpliesDual} implies dual stochastic boundedness.

Proposition \ref{proposition:weakImpliesDual} is of considerable practical value because techniques already exist for identifying pairs of distributions $\hat{v}_{ij}$ that weakly stochastically bound pairs of sets.
Indeed, many uncertainty sets including $\epsilon$-contamination sets and total variation neighbourhoods are already known to be weakly stochastically bounded in the sense of Definition \ref{definition:WSB} (cf.~ \cite[Section IV.A]{Molloy2017c} and references therein including \cite{Veeravalli1994,Unnikrishnan2011a}).
Furthermore, \eqref{eq:pDSB1} and \eqref{eq:pDSB2} are more intuitive, and typically more easily verified, than \eqref{eq:DSB1} and \eqref{eq:DSB2} due to the pseudo-distance interpretation of relative entropy.
Intuitively, \eqref{eq:pDSB1} holds when the bounding distributions $\tilde{\nu}_j$ are closer (in a relative-entropy sense) to the first distributions $\hat{v}_{ij}^0$ from the pairs $\hat{v}_{ij}$ than to the second distributions $\hat{v}_{ij}^1$.
Similarly, \eqref{eq:pDSB2} holds when the distributions $(\tilde{\nu}_i,\tilde{\nu}_j)$ are no further apart (in a relative-entropy sense) than the distributions in the pairs $\hat{v}_{ij}$.
When the pairs $\hat{v}_{ij}$ weakly stochastically bound the uncertainty sets, \eqref{eq:pDSB1} and \eqref{eq:pDSB2} therefore suggest that the distributions $\tilde{\nu}$ will typically exist provided that the uncertainty sets are no closer together (in a relative-entropy sense) than any distributions of the same change-type.
Finally, \eqref{eq:pDSB1} and \eqref{eq:pDSB2} hold trivially when the pairs of sets $\left( \mathcal{P}_i, \mathcal{P}_j \right)$ are each weakly stochastically bounded by pairs $\hat{v}_{ij} = (\tilde{\nu}_i,\tilde{\nu}_j)$ that share common distributions for $0 \leq i \leq J$ and $1 \leq j \leq J$ with $i \neq j$ (such as in the case of a two-sided alternative where $J = 2$) since \eqref{eq:pDSB2} holds with equality and \eqref{eq:pDSB1} becomes
 \begin{align*}
 \begin{split}
 &\min_{0 \leq i \leq J} \min_{\substack{1 \leq j \leq J\\j \neq i}} D \left( \tilde{\nu}_j \left\| \tilde{\nu}_i \right. \right)\\
 &\quad=\min_{0 \leq i \leq J} \min_{\substack{1 \leq j \leq J\\j \neq i}} \inf_{\nu_j \in \mathcal{P}_j} \left[ D \left( \nu_j \left\| \tilde{\nu}_i \right. \right) - D \left( \nu_j \left\| \tilde{\nu}_j \right. \right) \right],
 \end{split}
\end{align*}
which holds due to \eqref{eq:WSB1}.



\subsection{Asymptotically Robust Quickest Change Diagnosis}

We now establish our main result by combining our misspecified results with dual stochastic boundedness, and by defining the set of pairs associated with the distributions $\tilde{\nu}$ as
\begin{align*}
    \tilde{\Upsilon}
    &\triangleq \{ v_{ij} = \tilde{v}_{ij} \triangleq (\tilde{\nu}_i, \tilde{\nu}_j) : 0 \leq i,j \leq J, \; i \neq j, \; j \neq 0 \}.
\end{align*}

\begin{theorem}
\label{theorem:asymptoticMinimaxRobust}
Consider the sets $\{ \mathcal{P}_i \subset \mathcal{P} : 0 \leq i \leq J\}$ and suppose that they are dually stochastically bounded by $\hat{\Upsilon}$ and $\tilde{\nu} = \{ \tilde{\nu}_i \in \mathcal{P}_i : 0 \leq i \leq J \}$ in the sense of Definition \ref{def:DSB}.
Then the MCUSUM algorithm $\hat{\eta}(\hat{\Upsilon})$ with $h = |\log \gamma|$ solves both the standard non-robust problem \eqref{eq:standard} specified with $\nu = \tilde{\nu}$, and the robust problem \eqref{eq:robust} as $\gamma \to \infty$.
\end{theorem}
\begin{IEEEproof}
Since $C_\gamma$ is a subset of $C_\gamma(\tilde{\nu})$ (due to $\tilde{\nu}_i \in \mathcal{P}_i$ for all $0 \leq i \leq J$) and since Lemmas 1 and 2 of \cite{Oskiper2002} (see also \cite[Section VI]{Oskiper2002}) give that the MCUSUM algorithm $\hat{\eta}(\tilde{\Upsilon})$ with $h = |\log \gamma|$ is an asymptotic solution to \eqref{eq:standard}, we have that
 \begin{align}
  \label{eq:proofBounds}
  \begin{split}
  \inf_{\eta \in C_\gamma} W(\eta, \tilde{\nu})
  &\geq \inf_{\eta \in C_\gamma \left( \tilde{\nu} \right)} W(\eta, \tilde{\nu})\\
  &\sim W( \hat{\eta} (\tilde{\Upsilon}), \tilde{\nu})\\
  &= (1 + o(1)) \dfrac{| \log \gamma|}{\min\limits_{0 \leq i \leq J} \min\limits_{\substack{1 \leq j \leq J\\j \neq i}} D \left( \tilde{\nu}_j \left\| \tilde{\nu}_i \right. \right)}\\
  &\geq (1 + o(1)) \dfrac{| \log \gamma|}{\Delta_{*}^{\tilde{\nu},\hat{v}}}\\
  &\geq W(\hat{\eta}(\hat{\Upsilon}),\tilde{\nu})
  \end{split}
 \end{align}
 as $\gamma \to \infty$ where the second and third lines follow from Lemmas 1 and 2 of \cite{Oskiper2002}, the fourth line follows from dual stochastic boundedness \eqref{eq:DSB2}, and the last inequality follows from Theorem \ref{theorem:delayUpperBound} by noting that \eqref{eq:DSB2} implies that $0 < \Delta_{*}^{\tilde{\nu},\bar{v}}$.
 Under dual stochastic boundedness, namely \eqref{eq:DSB3}, Theorem \ref{theorem:falseBound} implies that $\hat{\eta}(\hat{\Upsilon})$ with $h = |\log \gamma|$ belongs to the set $C_\gamma \subset C_\gamma (\tilde{\nu})$, and so the properties of the infimum imply that 
 \begin{align*}
     \inf_{\eta \in C_\gamma} W \left(\eta, \tilde{\nu} \right) 
     \leq W(\hat{\eta} (\hat{\Upsilon}),\tilde{\nu}).
 \end{align*}
 Thus, the inequalities in \eqref{eq:proofBounds} must hold with equality, namely,
 \begin{align}
    \label{eq:inf}
     \inf_{\eta \in C_\gamma} W \left(\eta, \tilde{\nu} \right) 
     \sim \inf_{\eta \in C_\gamma \left( \tilde{\nu} \right)} W(\eta, \tilde{\nu})
     \sim W(\hat{\eta} (\hat{\Upsilon}),\tilde{\nu})
 \end{align}
 as $\gamma \to \infty$, proving the asymptotic optimality of $\hat{\eta} (\hat{\Upsilon})$ under \eqref{eq:standard} with $\nu = \tilde{\nu}$.
 Now, \eqref{eq:proofBounds} holding with equality implies that
 \begin{align}
  \label{eq:sup}
  \begin{split}
    W(\hat{\eta} (\hat{\Upsilon}),\tilde{\nu})
    &= (1 + o(1)) \dfrac{|\log \gamma|}{ \Delta_*^{\tilde{\nu},\hat{v}}}\\
    &\geq (1 + o(1)) \dfrac{|\log \gamma|}{ \Delta_*^{\nu,\hat{v}}}\\
    &\geq W(\hat{\eta}(\hat{\Upsilon}), \nu)
  \end{split}
 \end{align}
 as $\gamma \to \infty$ for all $\nu \in \mathcal{Q}$ where the second line follows from dual stochastic boundedness \eqref{eq:DSB1}, and the third line follows from Theorem \ref{theorem:delayUpperBound} by noting that \eqref{eq:DSB1} and \eqref{eq:DSB2} imply that $0 < \Delta_*^{\nu,\hat{v}}$.
 From \eqref{eq:inf} and \eqref{eq:sup} we therefore have that
 \begin{align*}
    \sup_{\nu \in \mathcal{Q}} W ( \hat{\eta} (\hat{\Upsilon}), \nu )
    \sim W ( \hat{\eta} (\hat{\Upsilon}), \tilde{\nu} )
    \sim \inf_{\eta \in C_\gamma} W (\eta, \tilde{\nu} )
 \end{align*}
 as $\gamma \to \infty$ and so $( \hat{\eta}(\hat{\Upsilon}), \tilde{\nu})$ is an asymptotic saddle point for our robust problem \eqref{eq:robust}.
  The proof is completed by noting that saddle points are minimax solutions and so $\hat{\eta}(\hat{\Upsilon})$ with threshold $h = |\log \gamma|$ solves \eqref{eq:robust} as $\gamma \to \infty$.
\end{IEEEproof}

Theorem \ref{theorem:asymptoticMinimaxRobust} establishes that the distributions $\tilde{\nu}$ introduced via dual stochastic boundedness of the uncertainty sets $\{\mathcal{P}_j : 0 \leq j \leq J\}$ in Definition \ref{def:DSB} are asymptotically least favourable for our robust problem \eqref{eq:robust} since the asymptotically robust MCUSUM algorithm $\hat{\eta}(\hat{\Upsilon})$ is also asymptotically optimal for the non-robust problem \eqref{eq:standard} specified by these LFDs $\nu = \tilde{\nu}$.
The asymptotic optimality of the MCUSUM algorithm $\hat{\eta}(\hat{\Upsilon})$ under the non-robust problem \eqref{eq:standard} specified by $\nu = \tilde{\nu}$ is novel since the existing optimality results of \cite[Section VI]{Oskiper2002} establish only that the correctly specified MCUSUM algorithm $\hat{\eta}(\tilde{\Upsilon})$ is an asymptotic solution to \eqref{eq:standard} with $\nu = \tilde{\nu}$.
By inspecting the proof of Theorem \ref{theorem:asymptoticMinimaxRobust}, we see that the asymptotic optimality of the MCUSUM algorithm $\hat{\eta}(\hat{\Upsilon})$ under the non-robust problem \eqref{eq:standard} with $\nu = \tilde{\nu}$ holds due to dual stochastic boundedness, specifically \eqref{eq:DSB2}, implying that the maximum worst-average detection and isolation delay $W(\eta,\nu)$ of $\hat{\eta}(\hat{\Upsilon})$ is equivalent to that of the correctly specified MCUSUM algorithm $\hat{\eta}(\tilde{\Upsilon})$.

Despite the asymptotic optimality of the asymptotically robust algorithm $\hat{\eta}(\hat{\Upsilon})$ under the non-robust problem \eqref{eq:standard} when $\nu = \tilde{\nu}$, when $\nu \neq \tilde{\nu}$ it will be suboptimal.
In the following corollary, we shall exploit the asymptotic bound on the delay of misspecified MCUSUM algorithms established in Theorem \ref{theorem:delayUpperBound} to characterise the extra delay incurred by the asymptotically robust algorithm $\hat{\eta}(\hat{\Upsilon})$ compared to the asymptotically optimal algorithm $\hat{\eta}(\Upsilon^*)$ under the non-robust problem \eqref{eq:standard} with $\nu \neq \tilde{\nu}$.

\begin{corollary}
 \label{corollary:costOfRobustness}
 Consider the uncertainty sets $\{ \mathcal{P}_i \subset \mathcal{P} : 0 \leq i \leq J\}$ and suppose that they are dually stochastically bounded by the pairs from the set $\hat{\Upsilon}$ and the distributions $\tilde{\nu} = \{ \tilde{\nu}_i \in \mathcal{P}_i : 0 \leq i \leq J \}$ in the sense of Definition \ref{def:DSB}.
 Furthermore, consider the distributions $\nu = \{ \nu_i \in \mathcal{P}_i : 0 \leq i \leq J \}$ and the associated set of pairs $\Upsilon^*$.
 Then the MCUSUM algorithms $\hat{\eta}(\hat{\Upsilon})$ and $\hat{\eta}(\Upsilon^*)$ with $h = |\log \gamma|$ satisfy
 \begin{align*}
     \dfrac{W(\hat{\eta}(\hat{\Upsilon}), \nu)}{W(\hat{\eta}(\Upsilon^*), \nu)}
     &\leq (1 + o(1)) \left( \dfrac{\min\limits_{0 \leq i \leq J} \min\limits_{\substack{1 \leq j \leq J,\; j \neq i}} D \left( {\nu}_j \left\| {\nu}_i \right. \right)}{\Delta_*^{\nu,\hat{v}}} \right)
 \end{align*}
 as $\gamma \to \infty$.
\end{corollary}
\begin{IEEEproof}
Dual stochastic boundedness, \eqref{eq:DSB1} and \eqref{eq:DSB2}, implies that $0 < \Delta_*^{\nu,\hat{v}}$ so Theorem \ref{theorem:delayUpperBound} gives that \eqref{eq:delayUpperBound} holds as $\gamma \to \infty$ with $\Delta_*^{\nu,\bar{v}} = \Delta_*^{\nu,\hat{v}}$.
 The proof is completed by dividing \eqref{eq:delayUpperBound} by the asymptotic equality
 \begin{align*}
  W ( \hat{\eta} ( \Upsilon^* ), \nu ) 
  = (1 + o(1)) \left( \dfrac{\left| \log \gamma \right|}{ \min\limits_{0 \leq i \leq J} \min\limits_{\substack{1 \leq j \leq J,\; j \neq i}} D \left( \nu_j \left\| \nu_i \right. \right) } \right)
 \end{align*}
 given by Lemmas 1 and 2 of \cite{Oskiper2002} as $\gamma \to \infty$.
\end{IEEEproof}

Corollary \ref{corollary:costOfRobustness} bounds the maximum performance loss incurred by using an asymptotically robust algorithm (i.e., $\hat{\eta}(\hat{\Upsilon})$) compared to an asymptotically optimal algorithm (i.e., $\hat{\eta}(\Upsilon^*)$).

\subsection{Relationship to Quickest Change Detection}
When $J = 1$, our robust quickest change diagnosis problem \eqref{eq:robust} reduces to the robust Lorden quickest change detection problem of \cite{Molloy2017c}, our dual stochastic boundedness condition (Definition \ref{def:DSB}) is equivalent to the weak stochastic boundedness condition of Definition \ref{def:DSB}, and the asymptotically robust MCUSUM algorithm $\hat{\eta}(\hat{\Upsilon})$ reduces to the asymptotically robust CUSUM rule of \cite{Molloy2017c}.
Theorem \ref{theorem:asymptoticMinimaxRobust} with $J = 1$ is also equivalent to the robust Lorden quickest change detection result of \cite[Theorem 3]{Molloy2017c}.
The key significance of our robust and misspecified quickest change diagnosis results is that they generalise and extend the quickest change detection results (without isolation) of \cite{Molloy2017c} by handling the simultaneous detection and isolation of an unknown change from one of $J > 1$ possible post-change uncertainty sets.

\section{Simulation Example}
\label{sec:results}

In this section, we consider the problem of diagnosing an unknown change in mean of a Gaussian process $Y_k \in \mathbb{R}^2$.
There are $J = 2$ change-types, and the uncertainty sets are:
\begin{align*}
 \mathcal{P}_0
 &= \left\{ \nu_0 = \mathcal{N} \left( \varphi, I \right) : \varphi = [\varphi_1, \varphi_2]', \; - \infty < \varphi_1, \varphi_2 \leq 0\right\}\\
  \mathcal{P}_1
 &= \left\{ \nu_1 = \mathcal{N} \left( \varphi, I \right) : \varphi = [\varphi_1, \varphi_2]', \; 0.4 \leq \varphi_1, \varphi_2 \leq 0.8 \right\}\\
  \mathcal{P}_2
 &= \left\{ \nu_2 = \mathcal{N} \left( \varphi, I\right) : \varphi = [\varphi_1, \varphi_2]', \; 1.5 \leq \varphi_1, \varphi_2 \leq \infty \right\}
\end{align*}
where $\mathcal{N} \left(\varphi, \Sigma^2\right)$ denotes the multivariate Gaussian distribution with mean vector $\varphi = [\varphi_1, \varphi_2]' \in \mathbb{R}^2$ and covariance matrix $\Sigma^2 \in \mathbb{R}^{2 \times 2}$.
By following the procedure for showing weak stochastic boundedness in the sense of Definition \ref{definition:WSB} described in \cite[Eq. (23)-(24)]{Molloy2017c}, the pairs of sets $\left( \mathcal{P}_i, \mathcal{P}_j \right)$ can each be seen to be weakly stochastically bounded by the pairs of distributions $\hat{v}_{ij}$ for $0 \leq i \leq 2$ and $1 \leq j \leq 2$, $i \neq j$ where $\hat{v}_{01}^0 = \hat{v}_{02}^0 = \mathcal{N} \left(0, I\right)$, $\hat{v}_{01}^1 = \mathcal{N} \left(0.4, I\right)$, $\hat{v}_{02}^1 = \mathcal{N} \left(1.5, I\right)$, $\hat{v}_{12}^0 = \hat{v}_{21}^1 = \mathcal{N} \left(0.8, I\right)$ and $\hat{v}_{12}^1 = \hat{v}_{21}^0 = \mathcal{N} \left(1.5, I\right)$.
Here, we use scalars to denote vectors of appropriate dimensions with repeated elements.
As discussed after Proposition \ref{proposition:weakImpliesDual}, the distributions from the pairs $\hat{v}_{ij}$ are natural candidates for the distributions $\tilde{\nu} = \{ \tilde{\nu}_0, \tilde{\nu}_1, \tilde{\nu}_2 \}$.
By selecting $\tilde{\nu}_0 = \mathcal{N} \left(0, I\right)$, $\tilde{\nu}_1 = \mathcal{N} \left(0.4, I\right)$ and $\tilde{\nu}_2 = \mathcal{N} \left(1.5, I\right)$, we have that \eqref{eq:pDSB1} and \eqref{eq:pDSB2} hold (which can be verified using the closed form expression for the relative entropy between Gaussians, cf.~\cite[Example 4.1.10]{Nikiforov1993}).
By applying Proposition \ref{proposition:weakImpliesDual}, it follows that the sets $\{\mathcal{P}_0, \mathcal{P}_1, \mathcal{P}_2\}$ are dually stochastically bounded by $\hat{\Upsilon} = \{\hat{v}_{01},\hat{v}_{02},\hat{v}_{12},\hat{v}_{21}\}$ and $\tilde{\nu} = \{ \tilde{\nu}_0, \tilde{\nu}_1, \tilde{\nu}_2\}$.

We implemented our asymptotically robust algorithm $\hat{\eta}(\hat{\Upsilon})$ with the pairs from $\hat{\Upsilon}$ together with the window-limited GLR algorithm of \cite{Lai2000}.
The GLR algorithm has a window-length parameter $w$ to trade-off detection performance for computational efficiency since it lacks an efficient recursive form.
We selected $w = 50$ and $w = 100$.

\begin{figure}[!t]
 \centering
 \includegraphics[width=3.3in]{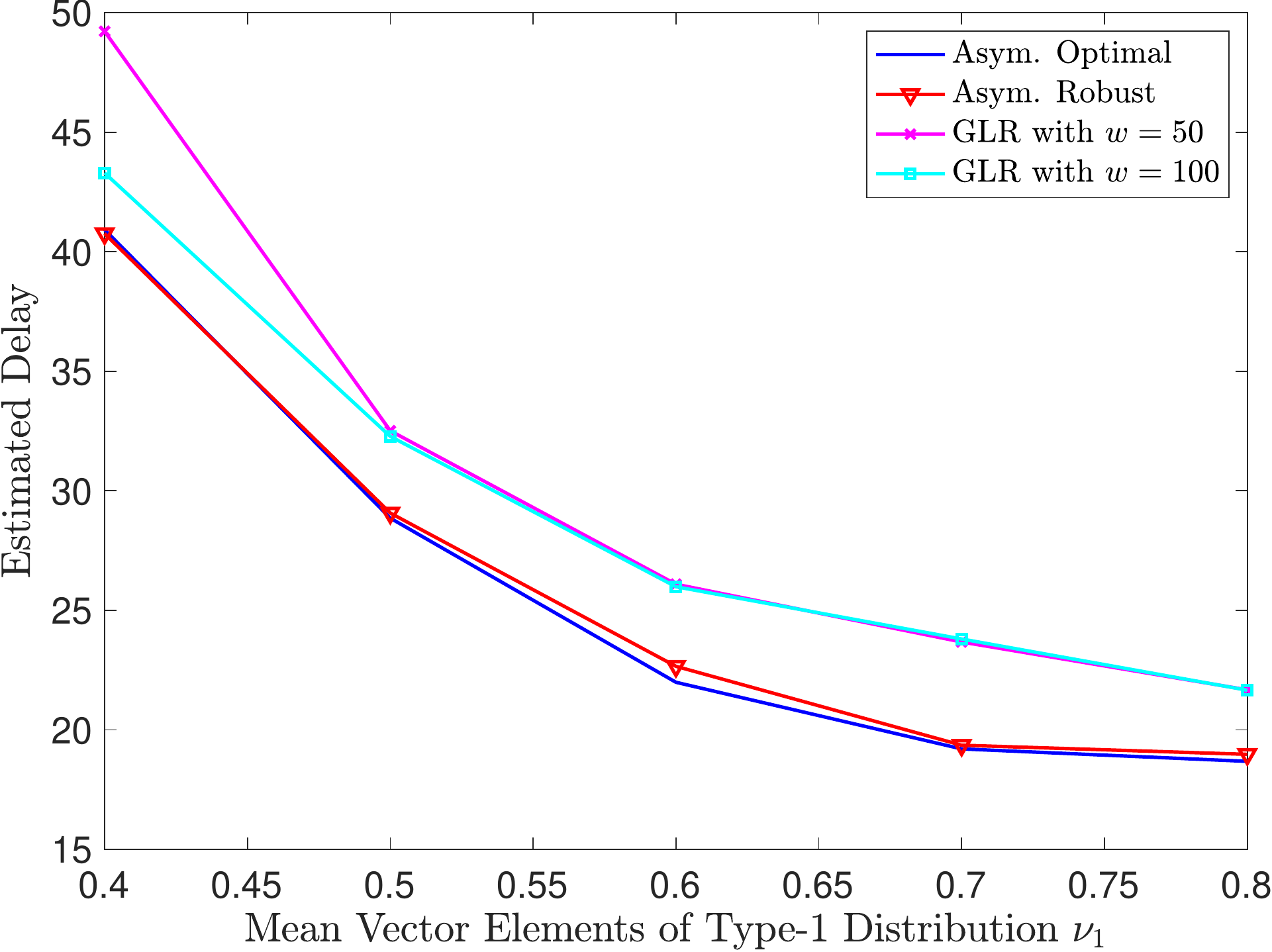}\\
    \small (a)\\[0.3cm]
    \includegraphics[width=3.3in]{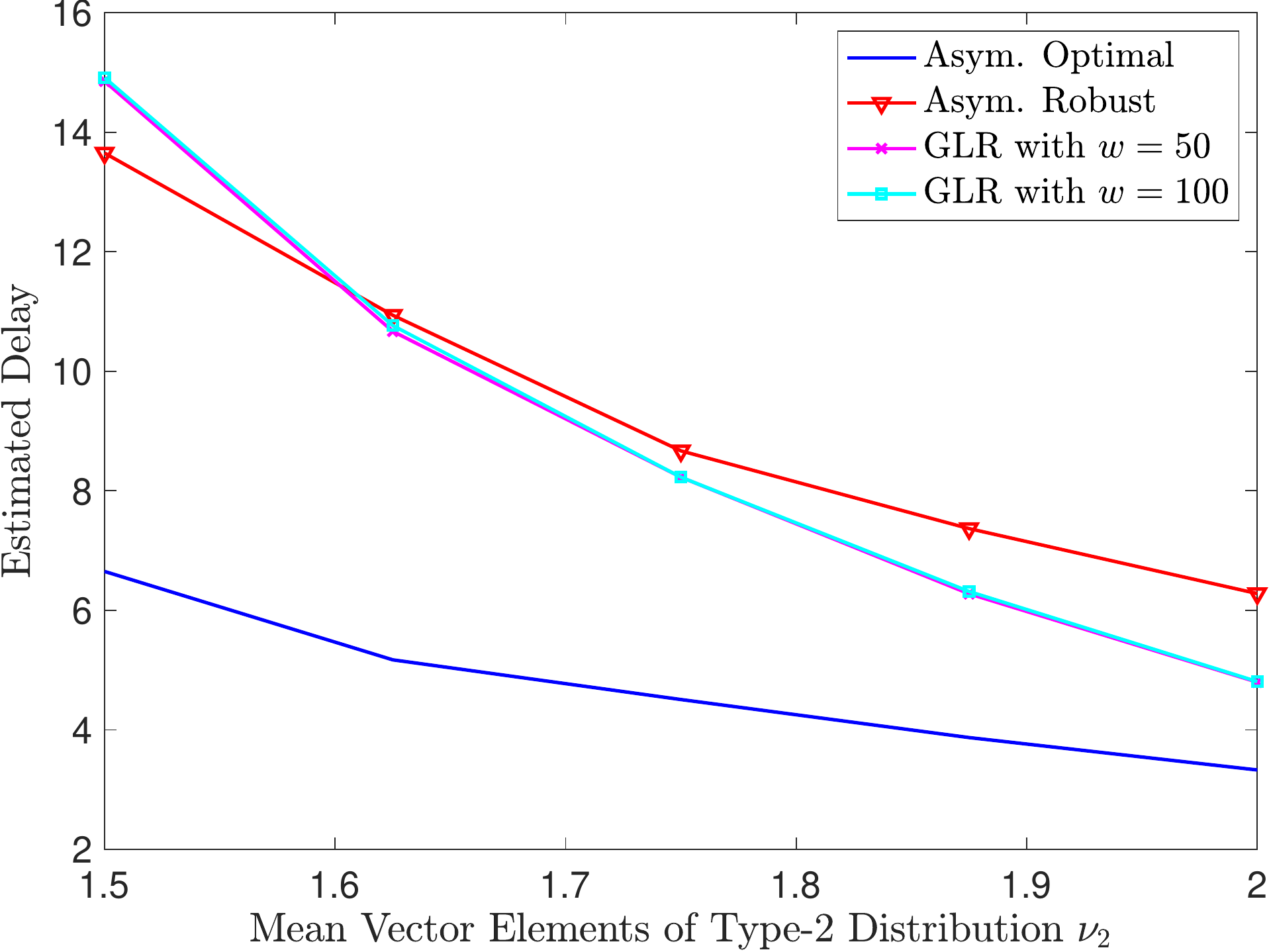}\\
    \small (b)
 \caption{Estimated detection and isolation delays against elements $\varphi_1 = \varphi_2$ of mean vector $\varphi$ corresponding to change-type (a) $j = 1$ and (b) $j = 2$ post-change distributions for an estimated mean time to false alarm or false isolation of $10000$.}
 \label{fig:delays}
\end{figure}

We simulated our asymptotically robust algorithm and the GLR algorithm on sequences with the true pre-change distribution $\nu_0 = \mathcal{N}(0,I)$ and with true (unknown) post-change distributions from either the type-$1$ or type-$2$ uncertainty sets $\mathcal{P}_1$ and $\mathcal{P}_2$.
We also implemented an asymptotically optimal MCUSUM algorithm $\hat{\eta}\left( \Upsilon^* \right)$ designed with unrealistic prior knowledge of the (unknown) true distributions $\nu$.
The detection and isolation delays $W(\eta,\nu)$ were estimated through the approximation of the expectations $E^{\nu_j}[\cdot]$ since the delays and expectation correspond for MCUSUM algorithms (by an argument similar to \cite[p.1380]{Moustakides1986}).
The estimated delays for type-$1$ changes with distributions $\nu_1 = \mathcal{N} \left( \varphi, I \right)$ where $0.4 \leq \varphi_1 = \varphi_2 \leq 0.8$ are reported in Fig.~\ref{fig:delays}(a) whilst those for type-$2$ changes with distributions $\nu_2 = \mathcal{N} \left( \varphi, I \right)$ where $1.5 \leq \varphi_1 = \varphi_2 \leq 2$ are reported in Fig.~\ref{fig:delays}(b).
We computed each delay from $500$ independent Monte Carlo runs, and selected algorithm thresholds so that the mean times to false alarm or false isolation were all $F\left( \eta, \nu\right) \approx 10000$.

From Fig.~\ref{fig:delays}, we see that all algorithms achieve their maximum (i.e., worst) delays at the bounding (or least favourable) distributions $\nu_1 = \tilde{\nu}_1 = \mathcal{N} (0.4,I)$ and $\nu_2 = \tilde{\nu}_2 = \mathcal{N} (1.5,I)$ for type-1 and type-2 changes, respectively.
Our asymptotically robust algorithm $\hat{\eta} ( \hat{\Upsilon} )$ corresponds to the asymptotically optimal algorithm $\hat{\eta}\left( \Upsilon^* \right)$ at $\nu_1 = \tilde{\nu}_1$, and therefore exhibits the (asymptotic) minimax property of (asymptotically) minimising the maximum delay over the distributions in $\{ \mathcal{P}_0, \mathcal{P}_1, \mathcal{P}_2\}$.
However, the performance of our asymptotically robust algorithm degrades (and differs from that of the asymptotically optimal algorithm) when the true distributions $\nu$ differ from the LFDs $\tilde{\nu}$.
In contrast, the performance of the GLR algorithm improves (and becomes closer to that of the asymptotically optimal algorithm) when the true distributions $\nu$ differ from the LFDs $\tilde{\nu}$.
The GLR algorithm is however especially sensitive to the choice of window length at the LFDs, and it is significantly more computationally complex than our (recursive) asymptotically robust algorithm.

\section{Conclusion}
\label{sec:conclusion}
We established new bounds on the performance of misspecified MCUSUM algorithms, and posed and solved a new robust quickest change diagnosis problem.
Our simulation results suggest that our asymptotically robust algorithm offers performance competitive with significantly more computationally complex GLR algorithms.
Future work will focus on relaxing the boundedness conditions under which our results are established, and formulating and solving robust versions of other quickest change diagnosis criteria (e.g., \cite{Nikiforov2003}).




\bibliographystyle{IEEEtran}

\bibliography{IEEEabrv,Library}

\end{document}